\documentclass{article}
\usepackage{amssymb}
\begin{document} 
\title{Symmetries of the equations of motion that are not shared by the Lagrangian}

\author{G.F.\ Torres del Castillo\\
Departamento de F\'isica Matem\'atica, Instituto de Ciencias \\
Universidad Aut\'onoma de Puebla, 72570 Puebla, Pue., M\'exico\\[1ex]
A.\ Moreno-Ruiz\\
Facultad de Ciencias F\'isico Matem\'aticas \\ Universidad Aut\'onoma de Puebla, 72570 Puebla, Pue., M\'exico}

\maketitle

\begin{abstract}
We show that if a Lagrangian is invariant under a transformation (with the invariance defined in the standard manner), then the equations of motion obtained from it maintain their form under the transformation. We also show that the converse is not true, giving examples of equations of motion that are form-invariant under a transformation, but these equations can be derived from a Lagrangian that is not invariant under such transformation. The conclusions are valid for discrete or continuous systems.
\end{abstract}



\section{Introduction}
The role of the symmetries is widely recognized in physics and, especially, in the theory of fields, to the extent that in many cases the symmetries are used as a guide to establish the basic equations of a theory or a model. For instance, in the framework of the special relativity, it is postulated that all the laws of physics must have the same form in all the inertial frames and, therefore, they have to maintain their form under the Lorentz transformations.

Usually, the textbooks on classical mechanics, quantum mechanics, or the theory of fields, contain some assertions without proof about the relation between the symmetries of the equations of motion, or the field equations, and the symmetries of the Lagrangians, or the Lagrangian densities, employed to obtain those equations.

On the one hand, it is vaguely claimed that if the Lagrangian is invariant under some transformation, then the equations of motion derived from it are form-invariant under the transformation (which may seem obvious, and not to require a proof). On the other hand, it is categorically claimed that in order to have a set of equations of motion form-invariant under some transformation, or family of transformations (e.g., the Lorentz group), the corresponding Lagrangian {\em must}\/ be invariant under that transformation or family of transformations (for a sample, see Refs.\ \cite{LL,Ba,HG,La,Ca,Gr,HF,Ja,FS,Ch,DT}).

The aim of this paper is to emphasize that the first claim is right, but the second one is wrong. Although these facts are given in the literature (see, e.g., Ref.\ \cite{PH}), it seems that they are not widely known.

In Sec.\ 2 we demonstrate that if a Lagrangian is invariant under a transformation (in the precise sense given below), then the equations obtained from such a Lagrangian are form-invariant under the transformation. We also give examples of differential equations with Lagrangians that are not invariant under all the transformations that leave invariant the form of the corresponding differential equation. In Sec.\ 3 we give analogous results for the case of fields.

\section{Systems with a finite number of degrees of freedom}
In order to simplify the discussion, in this section we start by considering in some detail the case of systems of ordinary differential equations, extending these results to the case of continuous systems or fields in Sec.\ 3.

\subsection{Invariance of the equations of motion}
A system of second-order ordinary differential equations
\begin{equation}
\ddot{q}_{i} = f_{i}(q_{j}, \dot{q}_{j}, t) \label{odes}
\end{equation}
($i, j = 1, 2, \ldots, n$), is invariant under the transformation
\begin{equation}
q'_{i} = q'_{i}(q_{j}, t), \qquad t' = t'(q_{j}, t) \label{ptr}
\end{equation}
if, making use of Eqs.\ (\ref{odes}), one obtains
\begin{equation}
\ddot{q}'_{i} = f_{i}(q'_{j}, \dot{q}'_{j}, t'), \label{podes}
\end{equation}
with the {\em same functions}\/ $f_{i}$ appearing in Eqs.\ (\ref{odes}), and the definitions
\begin{equation}
\dot{q}'_{j} \equiv \frac{{\rm d} q'_{j}}{{\rm d} t'}, \qquad \ddot{q}'_{j} \equiv \frac{{\rm d}^{2} q'_{j}}{{\rm d} t'^{2}}.
\label{pder}
\end{equation}

For instance, the differential equation
\begin{equation}
\ddot{q} = 0, \label{cacc}
\end{equation}
is invariant under the transformation
\begin{equation}
q' = \frac{q}{1 - sq}, \qquad t' = \frac{t}{1- sq}, \label{x8}
\end{equation}
for any fixed value of the parameter $s$, which is independent of the coordinates and the time. In fact, a straightforward computation yields
\begin{equation}
\dot{q}' = \frac{\dot{q}}{1 - sq + st \dot{q}} \label{pd}
\end{equation}
and
\begin{equation}
\ddot{q}' =  \left( \frac{1 - sq}{1 - sq + st \dot{q}} \right)^{3} \ddot{q}, \label{sd}
\end{equation}
hence, Eq.\ (\ref{cacc}) is equivalent to $\ddot{q}' = 0$, which has the same form as Eq.\ (\ref{cacc}).

Another simple example is given by the equation
\begin{equation}
\ddot{q} = - g, \label{caccn}
\end{equation}
where $g$ is a constant. This equation is invariant under the transformation \cite{AB}
\begin{equation}
q' = \frac{q}{1 - ts} - \frac{gt^{3}s}{2(1 - ts)^{2}}, \qquad t' = \frac{t}{1- ts}, \label{x8g}
\end{equation}
for any fixed value of the parameter $s$. Indeed,
\begin{equation}
\dot{q}' = (1 - ts) \dot{q} + qs + \frac{g}{2} \frac{t^{3} s^{2} - 3 t^{2} s}{1 - ts} \label{pdg}
\end{equation}
and
\begin{equation}
\ddot{q}' =  (1 - ts)^{3} (\ddot{q} +  g) - g, \label{sdg}
\end{equation}
thus, we see that Eq.\ (\ref{caccn}) is equivalent to $\ddot{q}' = - g$, which has the same form as Eq.\ (\ref{caccn}).

\subsection{Invariance of a Lagrangian}
Many systems of differential equations of the form (\ref{odes}) considered in physics can be expressed in the form
\begin{equation}
\frac{{\rm d}}{{\rm d} t} \frac{\partial L}{\partial \dot{q}_{i}} - \frac{\partial L}{\partial q_{i}} = 0, \label{le}
\end{equation}
in terms of an appropriate Lagrangian, $L(q_{i}, \dot{q}_{i}, t)$ (which is not unique), treating $(q_{i}, \dot{q}_{i}, t)$ as independent variables. However, as is well known, in the context of classical mechanics, a non-conservative system may not have a Lagrangian.

Since a single function, $L(q_{i}, \dot{q}_{i}, t)$, leads to the complete set of the differential equations of interest, it is natural to expect that the symmetries of $L$ be somehow related with the symmetries of the corresponding differential equations. What is needed is an appropriate definition of the invariance of $L$.

Even though it would seem natural to say that $L(q_{i}, \dot{q}_{i}, t)$ is invariant under the coordinate transformation (\ref{ptr}) if $L(q'_{i}, \dot{q}'_{i}, t') = L(q_{i}, \dot{q}_{i}, t)$, that definition is useful only if ${\rm d} t'/{\rm d} t = 1$, and, even in that case, it is highly restrictive (see below).

In order to find a convenient definition of the invariance of a Lagrangian under a coordinate transformation (\ref{ptr}), we begin by recalling that if the Lagrangian $L'$ is defined by
\begin{equation}
L'(q'_{i}, \dot{q}'_{i}, t') \equiv L \big( q_{j}(q'_{i}, t), \dot{q}_{j}(q'_{i}, \dot{q}'_{i}, t'), t(q'_{i}, t') \big) \frac{{\rm d} t}{{\rm d} t'}, \label{plag}
\end{equation}
then the Lagrange equations given by $L'$ are equivalent to those obtained from $L$. In fact, it can be shown that
\begin{equation}
\frac{{\rm d}}{{\rm d} t'} \frac{\partial L'}{\partial \dot{q}'_{i}} - \frac{\partial L'}{\partial q'_{i}} = \left( \frac{\partial q_{j}}{\partial q'_{i}} \frac{{\rm d} t}{{\rm d} t'} - \frac{\partial t}{\partial q'_{i}} \frac{{\rm d} q_{j}}{{\rm d} t'} \right) \left( \frac{{\rm d}}{{\rm d} t} \frac{\partial L}{\partial \dot{q}_{j}} - \frac{\partial L}{\partial q_{j}} \right) \label{ptle}
\end{equation}
(with sum over repeated indices), (see, e.g., Ref.\ \cite{EL}). Clearly, if the function $L'$ expressed in terms of the primed variables has the same form as the function $L$ expressed in terms of the unprimed variables, then the Lagrange equations for the primed variables have the same form as the Lagrange equations for the unprimed variables (consider, e.g., $L(x, \dot{x}, t) = \frac{1}{2} m \dot{x}^{2} - mgx$ and $L'(x', \dot{x}', t') = \frac{1}{2} m \dot{x}'^{2} - mgx'$; the corresponding Lagrange equations will have the same form).

Hence, the Lagrange equations given by a Lagrangian $L$ will be form-invariant under the transformation (\ref{ptr}) if
\[
L(q'_{i}, \dot{q}'_{i}, t') = L \big( q_{j}(q'_{i}, t), \dot{q}_{j}(q'_{i}, \dot{q}'_{i}, t'), t(q'_{i}, t') \big) \frac{{\rm d} t}{{\rm d} t'}
\]
[see Eq.\ (\ref{plag})] or, equivalently, interchanging the primed and unprimed variables, if
\begin{equation}
L \big( q'_{i}(q_{j}, t), \dot{q}'_{i}(q_{j}, \dot{q}_{j}, t), t'(q_{j}, t) \big) \frac{{\rm d} t'}{{\rm d} t} = L(q_{i}, \dot{q}_{i}, t).
\end{equation}
On the other hand, as is well known, the Lagrangians $L(q_{i}, \dot{q}_{i}, t)$ and
\[
L(q_{i}, \dot{q}_{i}, t) + \frac{\partial F}{\partial q_{i}} \dot{q}_{i} + \frac{\partial F}{\partial t}
\]
yield exactly the same Lagrange equations, for any function $F(q_{i}, t)$; thus, it is also true that the Lagrange equations given by a Lagrangian $L$ will be form-invariant under the transformation (\ref{ptr}) if
\begin{equation}
L \big( q'_{i}(q_{j}, t), \dot{q}'_{i}(q_{j}, \dot{q}_{j}, t), t'(q_{j}, t) \big) \frac{{\rm d} t'}{{\rm d} t} = L(q_{i}, \dot{q}_{i}, t) + \frac{\partial F}{\partial q_{i}} \dot{q}_{i} + \frac{\partial F}{\partial t}, \label{symvar}
\end{equation}
for some function $F$ of $(q_{i}, t)$ only. (If condition (\ref{symvar}) holds we say that Eqs.\ (\ref{ptr}) define a variational symmetry of $L$. The variational symmetries are relevant because if a Lagrangian possesses a one-parameter family of variational symmetries, then there is an associated constant of motion. See, e.g., Refs.\ \cite{AB,EL}, and the references cited therein.)

For instance, the differential equation (\ref{caccn}) can be obtained from the Lagrangian
\begin{equation}
L(q, \dot{q}, t) = {\textstyle \frac{1}{2}} m \dot{q}^{2} - mgq. \label{unif}
\end{equation}
Making use of Eqs.\ (\ref{x8g}) and (\ref{pdg}) we obtain [see Eq.\ (\ref{symvar})]
\[
\left( {\textstyle \frac{1}{2}} m \dot{q}'^{2} - mgq' \right) \frac{{\rm d} t'}{{\rm d} t} = \left\{ {\textstyle \frac{1}{2}} m \left[ (1 - ts) \dot{q} + qs + \frac{g}{2} \frac{t^{3} s^{2} - 3 t^{2} s}{1 - ts} \right]^{2} - mg \left( \frac{q}{1 - ts} - \frac{gt^{3}s}{2(1 - ts)^{2}} \right) \right\} \frac{1}{(1 - ts)^{2}}.
\]
The right-hand side of the last equation can be written as
\[
{\textstyle \frac{1}{2}} m \dot{q}^{2} - mgq + A(q,t) \dot{q} + B(q,t),
\]
where
\[
A(q,t) = \frac{mqs}{1 - ts} + \frac{mg (t^{3} s^{2} - 3 t^{2} s)}{2 (1 - ts)^{2}}
\]
and
\begin{eqnarray*}
B(q,t) & = & \frac{mq^{2}s^{2}}{2 (1 - ts)^{2}} + \frac{mg^{2} (t^{3} s^{2} - 3 t^{2} s)^{2}}{8 (1 - ts)^{3}} \\
& & \mbox{} + \frac{mgqs (t^{3} s^{2} - 3 t^{2} s)}{2 (1 - ts)^{3}} + \frac{mgt^{3}s}{2(1 - ts)^{4}} \\
& & \mbox{} + \frac{mgq (-3ts + 3t^{2}s^{2} - t^{3}s^{3})}{(1 - ts)^{3}}.
\end{eqnarray*}
In order to show that Eq.\ (\ref{symvar}) holds and, therefore, that (\ref{x8g}) is a variational symmetry of the Lagrangian (\ref{unif}), we simply verify that
\[
\frac{\partial A}{\partial t} = \frac{\partial B}{\partial q},
\]
which implies the existence of a function $F(q, t)$ such that $A = \partial F/\partial q$, $B = \partial F/\partial t$. (Note that what is relevant in the definition (\ref{symvar}) is the existence of $F$, and its explicit expression is not needed here.)

By contrast, the standard Lagrangian for the differential equation (\ref{cacc}),
\begin{equation}
L(q, \dot{q}, t) = {\textstyle \frac{1}{2}} m \dot{q}^{2} \label{stdfree}
\end{equation}
is {\em not}\/ invariant under the transformations (\ref{x8}). In fact, the difference [see Eq.\ (\ref{symvar})]
\[
{\textstyle \frac{1}{2}} m \dot{q}'^{2} \, \frac{{\rm d} t'}{{\rm d} t} - {\textstyle \frac{1}{2}} m \dot{q}^{2} = \frac{m\dot{q}^{2}}{2} \frac{1 - (1 - sq + st \dot{q})(1 - sq)^{2}}{(1 - sq + st \dot{q})(1 - sq)^{2}}
\]
is not a linear function of $\dot{q}$ for $s \not= 0$.

An additional example is given by Eq.\ (\ref{cacc}), which also follows from the Lagrangian
\begin{equation}
L(q, \dot{q}, t) = \dot{q} \ln \dot{q} - \dot{q}. \label{ns}
\end{equation}
As is well known, Eq.\ (\ref{cacc}), which is the equation of motion for a free particle (with respect to an inertial frame), is invariant under the Galilean transformations
\begin{equation}
q' = q - Vt, \qquad t' = t, \label{gal}
\end{equation}
where $V$ is a constant [analogous to the parameter $s$ appearing in Eqs.\ (\ref{x8}) and (\ref{x8g})]. In fact, it can be readily seen that
\[
\dot{q}' = \dot{q} - V, \qquad \ddot{q}' = \ddot{q},
\]
so that Eq.\ (\ref{cacc}) is trivially form-invariant.

One can verify that the Lagrangian (\ref{ns}) is not invariant under the Galilean transformations by calculating the difference
\[
\dot{q}' \ln \dot{q}' - \dot{q}' - (\dot{q} \ln \dot{q} - \dot{q}) = \dot{q} \ln \left( \frac{\dot{q} - V}{\dot{q}} \right) - V \ln (\dot{q} - V) + V,
\]
which is not a linear function of $\dot{q}$ for $V \not= 0$.

\section{Continuous systems and fields}
In the case of continuous systems or fields, the basic equations are similar to those presented above. Specifically, Eqs.\ (\ref{le}) are replaced by
\begin{equation}
\frac{{\rm d}}{{\rm d} x^{\alpha}} \frac{\partial {\mathcal L}}{\partial \phi_{i, \alpha}} - \frac{\partial {\mathcal L}}{\partial \phi_{i}} = 0, \label{fe}
\end{equation}
where ${\mathcal L}$ is a function of the field variables, $\phi_{i}$ ($i = 1, 2, \ldots, n$), which in turn are functions of the $m$ coordinates $x^{\alpha}$ ($\alpha = 1, 2, \ldots, m$) [through the field equations (\ref{fe})], $\phi_{i, \alpha} = \partial \phi_{i}/\partial x^{\alpha}$, and in the Lagrange equations (\ref{fe}), $\phi_{i}, \phi_{i, \alpha}$, and $x^{\alpha}$ are independent variables. The derivatives ${\rm d}/{\rm d} x^{\alpha}$ take into account the explicit and the implicit dependence on $x^{\alpha}$.

The Lagrangian density ${\mathcal L}(\phi_{i}, \phi_{i, \alpha}, x^{\alpha})$ is invariant under the transformation
\begin{equation}
\phi'_{i} = \phi'_{i}(\phi_{j}, x^{\beta}), \qquad x'^{\alpha} = x'^{\alpha}(\phi_{j}, x^{\beta}) \label{trf}
\end{equation}
if there exist functions $F^{\alpha}(\phi_{j}, x^{\beta})$ such that
\begin{equation}
{\mathcal L}(\phi'_{i}, \phi'_{i, \alpha}, x'^{\alpha}) \left| \frac{\partial (x'^{\alpha})}{\partial (x^{\beta})} \right| =  {\mathcal L}(\phi_{i}, \phi_{i, \alpha}, x^{\alpha}) + \frac{\partial F^{\alpha}}{\partial \phi_{i}} \phi_{i, \alpha} + \frac{\partial F^{\alpha}}{\partial x^{\alpha}}, \label{svf}
\end{equation}
with $\phi'_{i, \alpha} \equiv \partial \phi'_{i}/\partial x'^{\alpha}$. As in the case of systems with a finite number of degrees of freedom, the definition (\ref{svf}) assures that the Lagrange equations obtained from the Lagrangian density ${\mathcal L}$ be form-invariant under the transformation (\ref{trf}), and with each one-parameter family of variational symmetries there is an associated conserved current.

One of the favorite examples in the Lagrangian formulation of fields is that of the Maxwell equations. With the appropriate ``transformation law'' for the electromagnetic field (which amounts to say that the components of the electric and magnetic fields are the components of a two-index tensor field), the Maxwell equations are form-invariant under the Lorentz transformations.

The standard Lagrangian density for the source-free electromagnetic field (in cgs units) is
\begin{equation}
\mathcal{L} = \kappa ({\bf E}^{2} - {\bf B}^{2}), \label{emfs}
\end{equation}
where $\kappa$ is a constant, provided that the electromagnetic potentials, ${\bf A}, \varphi$, are the field variables. Since ${\bf E}^{2} - {\bf B}^{2}$ is one of the basic invariants of the electromagnetic field under the Lorentz transformations, the Lagrangian density (\ref{emfs}) satisfies the condition (\ref{svf}) with $F^{\alpha} = 0$.

The source-free Maxwell equations can be also derived from the Lagrangian density \cite{AA}
\begin{equation}
\mathcal{L} = {\bf B} \cdot \frac{1}{c} \frac{\partial {\bf E}}{\partial t} - {\bf B} \cdot \nabla \times {\bf B} - {\bf E} \cdot \frac{1}{c} \frac{\partial {\bf B}}{\partial t} - {\bf E} \cdot \nabla \times {\bf E}, \label{emfns}
\end{equation}
where ${\bf E}$ and ${\bf B}$ are the field variables. Indeed, substituting (\ref{emfns}) into (\ref{fe}) one obtains the ``evolution equations''
\begin{equation}
\frac{1}{c} \frac{\partial {\bf E}}{\partial t} = \nabla \times {\bf B}, \qquad \frac{1}{c} \frac{\partial {\bf B}}{\partial t} = - \nabla \times {\bf E}. \label{evo}
\end{equation}
The remaining Maxwell equations ($\nabla \cdot {\bf B} = 0$ and $\nabla \cdot {\bf E} = 0$) are imposed as initial conditions. One can see that if we consider an electromagnetic field such that $\nabla \cdot {\bf B} = 0$ and $\nabla \cdot {\bf E} = 0$ at some initial time, then the evolution equations (\ref{evo}) guarantee that they will hold all the time. (In this sense, the situation is analogous to that in the standard Lagrangian density (\ref{emfs}), where, by virtue of the postulated relations ${\bf B} = \nabla \times {\bf A}$ and ${\bf E} = - \nabla \varphi - (1/c) \, \partial {\bf A}/\partial t$, two of the Maxwell equations are identically satisfied. One advantage of (\ref{emfns}) is that it is gauge-independent.)

The interest of the Lagrangian density (\ref{emfns}) here is that it is not invariant under the Lorentz transformations. (This Lagrangian density is also interesting because it shows that, in the case of the source-free Maxwell equations, the idea that the electromagnetic potentials have to be used as the field variables is wrong.) In order to show that (\ref{emfns}) is not invariant under the Lorentz transformations, it will be convenient to consider an ``infinitesimal'' boost. Recalling that under a boost in an arbitrary direction, characterized by the constant vector \mbox{\boldmath $\beta$}, the electromagnetic field transforms according to
\begin{eqnarray*}
{\bf E}' & = & \gamma ({\bf E} + \mbox{\boldmath $\beta$} \times {\bf B}) - \frac{\gamma^{2}}{\gamma + 1} (\mbox{\boldmath $\beta$} \cdot {\bf E}) \mbox{\boldmath $\beta$}, \\
{\bf B}' & = & \gamma ({\bf B} - \mbox{\boldmath $\beta$} \times {\bf E}) - \frac{\gamma^{2}}{\gamma + 1} (\mbox{\boldmath $\beta$} \cdot {\bf B}) \mbox{\boldmath $\beta$},
\end{eqnarray*}
where, as usual, $\gamma = (1 - \mbox{\boldmath $\beta$}^{2})^{-1/2}$, we find that, to first order in \mbox{\boldmath $\beta$},
\[
\delta {\bf E} \equiv {\bf E}' - {\bf E} = \mbox{\boldmath $\beta$} \times {\bf B}, \qquad \delta {\bf B} \equiv {\bf B}' - {\bf B} = - \mbox{\boldmath $\beta$} \times {\bf E},
\]
and, therefore (taking into account the fact that the Jacobian of a proper Lorentz transformation is equal to 1),
\[
\delta \mathcal{L} \equiv \mathcal{L}(\phi'_{i}, \phi'_{i, \alpha}, x'^{\alpha}) - \mathcal{L}(\phi_{i}, \phi_{i, \alpha}, x^{\alpha})
\]
is, to first order in \mbox{\boldmath $\beta$},
\begin{eqnarray*}
\delta \mathcal{L} & = & - \mbox{\boldmath $\beta$} \times {\bf E} \cdot \left( \frac{1}{c} \frac{\partial {\bf E}}{\partial t} - \nabla \times {\bf B} \right) + {\bf B} \cdot (\mbox{\boldmath $\beta$} \nabla \cdot {\bf E}) \\
& & - \mbox{\boldmath $\beta$} \times {\bf B} \cdot \left( \frac{1}{c} \frac{\partial {\bf B}}{\partial t} + \nabla \times {\bf E} \right) - {\bf E} \cdot (\mbox{\boldmath $\beta$} \nabla \cdot {\bf B}),
\end{eqnarray*}
which is linear in the derivatives of the field variables, but if $\delta \mathcal{L}$ were of the form
\[
\frac{\partial F^{\alpha}}{\partial \phi_{i}} \phi_{i, \alpha} + \frac{\partial F^{\alpha}}{\partial x^{\alpha}}
\]
[see Eq.\ (\ref{svf})], centering the attention to the coefficients of $E_{i, 0} \equiv (1/c) \, \partial E_{i}/\partial t$, we would have
\[
\frac{\partial F^{0}}{\partial E_{i}} = \varepsilon_{ijk} E_{j} \beta_{k},
\]
where the $E_{i}, \beta_{i}$ are Cartesian components of ${\bf E}$ and \mbox{\boldmath $\beta$}, respectively, and $\varepsilon_{ijk}$ is the Levi-Civita symbol. We get a contradiction, because then
\[
\frac{\partial^{2} F^{0}}{\partial E_{l} \partial E_{i}} = \varepsilon_{ilk} \beta_{k},
\]
the left-hand side must be symmetric in the indices $l$ and $i$, while the right-hand side is skewsymmetric in those indices.

\section{Concluding remarks}
In view of the results presented above, the following question arises: Given an equation or a set of equations form-invariant under a given transformation, is it always possible to find a Lagrangian (or Lagrangian density) invariant under that transformation? (see, e.g., Ref.\ \cite{Le}). In fact, it is known that there exist sets of ordinary differential equations which cannot have a Lagrangian (see, e.g., Ref.\ \cite{Do}). This point is important because, in the current research, the attention is usually focused on the Lagrangians rather than on the field equations themselves.

\section*{Acknowledgment}
One of the authors (A.M.R.) wishes to thank the Vicerrector\'{\i}a de Investigaci\'on y Estudios de Posgrado of the Universidad Aut\'onoma de Puebla for financial support.

\end{document}